\documentclass[pre,twocolumn,showpacs]{revtex4}
\usepackage{graphicx,amsmath,amssymb}

\newcommand{\figwidth}{0.90\columnwidth}
\newcommand{\eq}[1]{Eq.(\ref{#1})}
\newcommand{\fig}[1]{Fig.~\ref{#1}}
\newcommand{\sect}[1]{Sec.~\ref{#1}}
\newcommand{\avg}[1]{{\langle #1 \rangle}}
\newcommand{\olcite}[1]{Ref.~\onlinecite{#1}}

\newcommand{\kb}{ k_{\rm B} }
\newcommand{\rhocp}{ \rho_{\rm cp} }
\newcommand{\riso}{ \rho^\star_{\rm ISO} }
\newcommand{\rnem}{ \rho^\star_{\rm NEM} }

\newcommand{\Pn}{ P(N) }
\newcommand{\gin}{ \gamma_{\rm IN} }

\newcommand{\lx}{ L_{\rm x} }
\newcommand{\ly}{ L_{\rm y} }
\newcommand{\lz}{ L_{\rm z} }

\begin{document}

\title{Isotropic-nematic interfacial tension of hard and soft rods:  
application of advanced grand canonical biased sampling techniques}

\author{R. L. C. Vink, S. Wolfsheimer and T. Schilling}

\affiliation{Institut f\"ur Physik, Johannes Gutenberg-Universit\"at,
D-55099 Mainz, Staudinger Weg 7, Germany}

\date{\today}

\begin{abstract} 

Coexistence between the isotropic and the nematic phase in suspensions of
rods is studied using grand canonical Monte Carlo simulations with a bias
on the nematic order parameter. The biasing scheme makes it possible to
estimate the interfacial tension $\gin$ in systems of hard and soft rods.  
For hard rods with $L/D=15$, we obtain $\gin \approx 1.4 \, \kb T/L^2$,
with $L$ the rod length, $D$ the rod diameter, $T$ the temperature, and
$\kb$ the Boltzmann constant. This estimate is in good agreement with
theoretical predictions, and the order of magnitude is consistent with
experiments.

\end{abstract}



\maketitle

\section{Introduction}

The aim of this paper is to present a computation of the interfacial
tension $\gin$ between the coexisting isotropic and nematic (IN) phase in
suspensions of monodisperse hard rods via computer simulation. While the
hard-rod fluid simplifies experimental reality, ignoring for example
long-ranged interactions and polydispersity \cite{chen.gray:2002}, it
nevertheless captures the main mechanism of the IN phase transition and
serves as a valuable model system. Experiments have shown that $\gin$ is
very small, typically in the range 10$^{-3}$--10$^{-4}$~mN/m
\cite{chen.gray:2002}, which makes it difficult to extract $\gin$ from
simulation data. Simulation estimates of $\gin$ are therefore rare, and
have only been reported for ellipsoids \cite{mcdonald:2000,
akino.schmid.ea:2001}, soft rods \cite{vink.schilling:2005}, and lattice
models \cite{cleaver.allen:1993}. Theoretical estimates are more abundant
\cite{chen.noolandi:1992, koch.harlen:1999, schoot:1999,
velasco.mederos:2002, allen:2000*b, shundyak.roij:2001}, but are usually
obtained in the Onsager limit \cite{onsager:1949} of infinite rod length
($L/D \to \infty$). The case of finite rod length is more difficult to
describe theoretically, but has been addressed in
\olcite{velasco.mederos:2002} using density functional theory, and in
\olcite{schoot:1999} using a scaling relation. At the time of writing, no
simulation estimate of $\gin$ for the hard-rod fluid has been reported.
Such an estimate would clearly be valuable to test theoretical
predictions, and to see if the order of magnitude of $\gin$ observed in
experiments is reproduced.

Despite its simplicity, simulating the hard-rod fluid is not trivial
\cite{bolhuis.frenkel:1997, dijkstra.roij.ea:2001}. The bottleneck is the
hard particle interaction, which complicates both molecular dynamics (MD)
and Monte Carlo (MC) methods. In the case of MD, the discontinuous
potential prevents the calculation of smooth forces. In the case of MC,
equilibration times are long due to very low acceptance rates. An
important improvement is the use of soft interactions, as was done for
ellipsoids \cite{mcdonald:2000, akino.schmid.ea:2001} and rods
\cite{al-barwani.allen:2000, vink.schilling:2005}. By using soft
interactions, the qualitative phase behavior is usually retained, but
simulations become much more efficient. Moreover, MC simulations in the
grand canonical ensemble become possible, enabling the investigation of IN
coexistence via the probability distribution in the particle number
density. This technique is well established in simulations of fluid-vapor
coexistence \cite{potoff.panagiotopoulos:2000, gozdz:2003,
virnau.muller.ea:2004, muller.macdowell:2000, vink.horbach:2004*1}, and
was recently extended to IN coexistence in suspensions of soft rods
\cite{vink.schilling:2005}. The advantage of grand canonical simulations
is that the coexistence densities, as well as the interfacial tension, can
be obtained.

Since coexisting phases are separated by a free energy barrier arising
from the interfacial tension \cite{wilding:2001}, it is essential to use a
biased sampling scheme to access regions of high free energy. In
simulations of fluid-vapor coexistence, the bias is usually put on the
density. While a density bias has also been used to simulate IN
coexistence \cite{vink.schilling:2005}, this choice is not optimal. In
simulations that rely on standard MC moves, such as random translations
and rotations of single particles, it is difficult to reach the nematic
phase starting in the isotropic phase simply by increasing the density
because the orientational degrees of freedom relax only very slowly
\cite{williams.philipse:2003}. This effect is called ``jamming'', and it
explains why the simulations of \olcite{vink.schilling:2005} were limited
to rather small systems.

In this work, grand canonical MC simulations using a bias on the nematic
order parameter are performed. As we will show, this approach is much less
susceptible to jamming, and enables simulations of large systems. This in
turn allows for accurate estimates of the interfacial tension in
suspensions of soft rods. As an additional bonus, a bias on the nematic
order parameter paves the way towards grand canonical simulations of hard
rods, enabling a simulation estimate of $\gin$ for the hard-rod fluid.

The outline of this paper is as follows: First, we introduce the liquid
crystal model used in this work. The biased sampling scheme is described
next. The results are presented in \sect{sec:res}. We end with a summary 
and comparison to theoretical predictions in the last section.

\section{Model and order parameters}

We consider rods of elongation $L$ and diameter $D$. The simulations are
performed in a three dimensional box of size $\lx \times \ly \times \lz$
using periodic boundary conditions in all dimensions. In this work, we fix
$\lx=\ly$, but we allow for elongation in the remaining dimension $\lz
\geq \lx$. Moreover, to avoid double interactions between rods through the
periodic boundaries, we set $\lx > 2 \, L$. The position of the center of
mass of rod $i$ is denoted $\vec{r}_i$, and its orientation $\vec{u}_i$,
with normalization $|\vec{u}_i|=1$. The interaction between two rods $i$
and $j$ is given by a pair potential of the form
\begin{eqnarray}
\label{eq:pot}
  v_{ij} (r) &=&
  \begin{cases}
  \epsilon & r < D, \\
  0 & {\rm otherwise},
  \end{cases}
\end{eqnarray}
with $r$ the distance between two line segments of length $L$, see also
\olcite{vink.schilling:2005}. The total energy is thus a function of the center
of mass coordinates and the orientations of all rods
\begin{equation}
\label{eq:tot}
  E(\vec{r}_1, \ldots, \vec{r}_N; \vec{u}_1, \ldots, \vec{u}_N) 
  = \sum_{i=1}^N \sum_{j=i+1}^N v_{ij},
\end{equation} 
with $N$ the number of rods in the system (in the following we will drop the
$\vec{r}_i$ and $\vec{u}_i$ dependences in our notation).

To investigate the IN transition, the density and the average rod alignment are
used as order parameters. Since the density in the isotropic phase is lower than
in the nematic phase, the rod number density $\rho=N/V$ may be used to
distinguish the phases, with $V$ the volume of the simulation box. Following
convention, we also introduce the reduced density $\rho^\star = \rho / \rhocp$,
with $\rhocp = 2 / [\sqrt{2} + (L/D)\sqrt{3}]$ the density of regular close
packing of hard rods \cite{bolhuis.frenkel:1997}.

In the nematic phase the rods are on average aligned, whereas in the
isotropic phase the rods are randomly oriented. Therefore, the nematic
order parameter may also be used to distinguish the phases. The latter
quantity is defined in terms of the orientational tensor $Q$, whose
components are given by
\begin{equation}
\label{eq:s2}
 Q_{\alpha\beta} = \frac{1}{2 N} \sum_{i=1}^N
   \left( 3 u_{i\alpha} u_{i\beta} - \delta_{\alpha\beta} \right),
\end{equation}
with $u_{i\alpha}$ the $\alpha$ component ($\alpha = x,y,z$) of the orientation
of rod $i$, and $\delta_{\alpha\beta}$ the Kronecker delta. In this work, the
maximum eigenvalue $S$ of the orientational tensor is taken as nematic order
parameter, being close to unity in the nematic phase, and close to zero in the
isotropic phase. The eigenvector corresponding to $S$ is called the director,
and it measures the preferred direction of the rods in the nematic phase.

\section{Simulation method}

We study IN coexistence via grand canonical MC simulations. In the grand
canonical ensemble, the volume, the temperature $T$, and the chemical
potential $\mu$ are fixed, while the number of rods in the simulation box
fluctuates. Insertion and removal of rods are attempted with equal
probability and accepted with appropriate Metropolis rules to be given
later. The aim of grand canonical simulations is to measure the
probability distribution in the number of particles $\Pn$. At the
coexistence chemical potential, $\Pn$ becomes bimodal with two peaks of
equal area. An example distribution is shown in \fig{compare}, where we
have plotted the logarithm of $P(N)$. The peak locations yield the
coexistence densities; the average height of the peaks $\Delta \Omega$ in
$\kb T \ln \Pn$ is the free energy barrier separating the phases, with
$\kb$ the Boltzmann constant. In three dimensions using periodic boundary
conditions and for sufficiently large systems, the barrier is related to
the interfacial tension via $\gin = \Delta\Omega / (2\lx^2)$, where $\lx$
is the lateral dimension of the simulation box \cite{binder:1982}.

In simulations, the free energy barrier presents a problem. Unless
$\Delta\Omega$ is small, such as close to a critical point, simulations
rarely cross the barrier, and spend most time in only one of the two
phases.  Biased sampling techniques are required to overcome the barrier.
In general, these techniques aim to construct a weight function $W(\zeta)$
of some bias variable $\zeta$. The weight function is constructed such
that a simulation using a modified potential $E'(\zeta) = E + \kb T \,
W(\zeta)$ yields a uniform probability distribution in the bias variable,
with $E$ the potential of the original system. The grand canonical
acceptance rules using the modified potential read as
\begin{eqnarray}
  \label{eq:ins}
  A(N,\zeta_0 \to N+1,\zeta_1) = \hspace{3cm} \nonumber \\
    \min \left[1, \frac{V}{N+1}
    e^{ -\beta (\Delta E - \mu) - W(\zeta_1) + W(\zeta_0) } \right], \\
  \label{eq:rem}
  A(N,\zeta_0 \to N-1,\zeta_1) = \hspace{3cm} \nonumber \\
    \min \left[1, \frac{N}{V}
    e^{ -\beta (\Delta E + \mu) + W(\zeta_1) - W(\zeta_0) } \right],
\end{eqnarray}
for the insertion and removal of a single particle, respectively
\cite{landau.binder:2000, frenkel.smit:2001}. Here, $\zeta_0$ and
$\zeta_1$ denote, respectively, the value of the bias parameter in the
initial and final state, $\Delta E$ is the potential energy difference
between initial and final state given by \eq{eq:tot}, and $\beta=1/(\kb
T)$. For a properly constructed $W(\zeta)$, the biased simulation samples
all states $\zeta$ with uniform probability. Once $W(\zeta)$ is known, the
distribution $\Pn$ of the unbiased system can be constructed.

One is rather free in choosing the bias variable. The best choices are
variables that change significantly when going from one phase to the
other. For fluid-vapor transitions, a natural bias is the particle number
density. In the case of IN coexistence, the density is still a valid
variable because of the density gap between the isotropic and the nematic
phase. This was used in \olcite{vink.schilling:2005} to study IN
coexistence in suspensions of soft rods. Whether a bias on density in
systems of elongated particles is efficient, depends on how easily a dense
isotropic phase can rearrange itself to become nematic. In practice, the
jamming effect limits density biased sampling to rather small systems and
soft interactions. As it turns out, for IN transitions, a much more
powerful bias variable is the nematic order parameter $S$. Note, however,
that phase coexistence is defined in terms of $\Pn$. Therefore, in a
simulation which biases on $S$, the distribution $\Pn$ must still be
reconstructed. To this end, histograms in both the particle number $N$, as
well as in $S$, have to be measured. In this section, we explain how the
bias on $S$, and the subsequent reweighting in $N$ and $S$, are
implemented. It is convenient to discuss the more straightforward
procedure of a density bias first.
	
\subsection{Biased sampling on $\rho$}

A convenient method to bias on density is {\it Successive Umbrella
Sampling} (SUS) \cite{virnau.muller:2004}. Here we describe the algorithm
in its simplest form; refinements are given in the original reference. The
choice of the sampling algorithm is not crucial. The general principles
also apply to other schemes, such as conventional umbrella sampling
\cite{torrie.valleau:1977}, multicanonical sampling
\cite{berg.neuhaus:1992}, Wang-Landau sampling \cite{wang.landau:2001}, or
hyperparallel tempering \cite{yan.de-pablo:2000}.

The aim is to construct a function $W(N)$ of the number of particles such
that a simulation using the modified potential $E'(N) = E + \kb T \, W(N)$
yields a uniform distribution in $N$, with $E$ given by \eq{eq:tot}. The
modified potential thus contains an explicit dependence on the bias
variable $N$. Following \olcite{virnau.muller:2004}, the particle number
axis is divided into equally sized intervals called windows, starting with
some minimum number of particles $N_0$. In the first window, the number of
particles is confined to $N_0 \leq N \leq N_0+1$, in the second window to
$N_0+1 \leq N \leq N_0+2$, and in the $i$-th window to $N_0+i-1 \leq N
\leq N_0+i$. In this example, the window size equals a single particle but
this choice is not essential: SUS works just as well using larger windows
\cite{virnau.muller:2004}. The choice of the window size is not completely
arbitrary. Choosing the windows too large leads to poor sampling
statistics at the window boundaries; choosing the windows too small runs
the risk that certain relaxation pathways are cut-off. In practice, a
compromise needs to be made.

The idea of SUS is to construct $W(N)$ by simulating the windows
separately and successively. Starting in the first window ($i=1$), grand
canonical MC moves are attempted (optionally combined with canonical moves
such as translations and rotations), with the constraint that states
outside the window bounds are rejected to fulfill detailed balance at the
window boundaries. The relevant weights in the first window are $W(N_0)$
and $W(N_0+1)$, which we initially set to zero.  We then record $f_{\rm
L}^1$ and $f_{\rm H}^1$, counting the occurrence of the state with $N_0$
and $N_0+1$ particles, respectively. In this notation, the subscripts
``L'' and ``H'' refer to the ``lower'' and ``higher'' window bound,
respectively, while the superscript refers to the window number. To obtain
a uniform distribution in $N$, the ratio of the counts should be unity.  
This will generally not be the case, but is enforced by updating the
weight of the higher window bound to
\begin{equation}
\label{eq:update}
  W_{\rm new} (N_0+i) = W_{\rm old} (N_0+i) + 
    \ln (f_{\rm H}^i / f_{\rm L}^i),
\end{equation}
leaving the weight of the lower bound $W(N_0+i-1)$ unchanged, where $i$ is
the window number. In case $f_{\rm H}^i > f_{\rm L}^i$, the effect of this
modification is a lower insertion rate, see \eq{eq:ins}, and a higher
removal rate, see \eq{eq:rem}, leading to a count ratio closer to one. The
latter can be checked by performing a second simulation using the updated
weight (the reasoning for $f_{\rm H}^i < f_{\rm L}^i$ is similar). In
practice, it may occur that one of the counts is zero. It is then
necessary to modify $W(N_0+i)$ by hand first, before starting the
simulation. Note also that long simulation runs may be required to obtain
the count ratio accurately.

Having simulated the first window, $W(N_0)$ and $W(N_0+1)$ are known. The
choice $W(N_0)=0$ is arbitrary but has no physical consequences since it
merely shifts the potential by a constant. Next, we consider window~2,
where the number of particles is allowed to fluctuate between $N_0+1$ and
$N_0+2$, with respective weights $W(N_0+1)$ and $W(N_0+2)$. An important
optimization of \olcite{virnau.muller:2004} is to linearly extrapolate the
known weights $W(N_0)$ and $W(N_0+1)$ to obtain an estimate for $W(N_0+2)$
(note that for the third and subsequent windows, quadratic extrapolation
can be used). The simulations in the second window are then performed
using the extrapolated estimate, and the respective counts, $f_{\rm L}^2$
and $f_{\rm H}^2$, of visiting the state with $N_0+1$ and $N_0+2$
particles, are recorded. Finally, the weight $W(N_0+2)$ is updated using
\eq{eq:update}, leaving the other weight $W(N_0+1)$ unchanged, and the
next window is considered.

The above procedure is repeated until all windows of interest have been
simulated, and the corresponding weight function $W(N)$ is constructed.
The sought-for distribution in the number of particles $\Pn$ is trivially
obtained via $\Pn = C e^{W(N)}$, with $C$ a normalization constant
\cite{wilding:2001}.

\subsection{Biased sampling on $S$}
\label{sec:s2}

Next, we consider the extension to a bias on the nematic order parameter
$S$. Here, the modified potential reads as $E'(S) = E + \kb T \, W(S)$,
with $E$ given by \eq{eq:tot}. The aim is to construct $W(S)$ such that a
simulation using the modified potential samples all values of $S$ with 
uniform probability.

\begin{figure}
\begin{center}
\includegraphics[clip=,width=\figwidth]{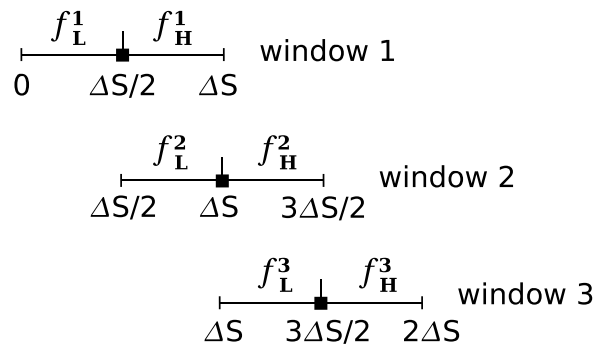}

\caption{\label{umb} Schematic illustration of biased sampling on the
nematic order parameter. See details in text.}
 
\end{center}
\end{figure} 

The windows are obtained by dividing the nematic order parameter axis into
equally sized intervals of width $\Delta S$. In the first window, the nematic
order parameter is confined to $0 \leq S < \Delta S$, in the second window to
$\Delta S/2 \leq S < 3\Delta S/2$, and in the $i$-th window to $(i-1)\Delta S/2
\leq S < (i+1) \Delta S/2$, see \fig{umb}. The windows thus partially overlap.
To sample both the isotropic and the nematic phase, the sampling range should
span from $S=0$ to $S \approx 1$. Note that $S$ is a continuous variable,
whereas the density (expressed in the number of particles) is discrete.
Therefore, a natural width for the windows does not exist, and one is forced to
choose $\Delta S$ rather arbitrarily. We found that $\Delta S \approx
0.001-0.002$ gives good results, which means that ${\cal O}(10^3)$ windows are
required to sample the transition. A consequence of discretizing the nematic
order parameter is that $W(S)$ is defined in steps of $\Delta S/2$. Therefore,
in the $i$-th window, $W(S)$ assumes only two distinct values
\begin{eqnarray*}
  W (S) &=&
  \begin{cases}
    W_{i-1} & (i-1)\Delta S/2 \leq S < S_{\rm M} \\
    W_i &  S_{\rm M} \leq S < (i+1) \Delta S/2,
  \end{cases}
\end{eqnarray*}
with $S_{\rm M} = i \Delta S/2$ the center of the window (note that $i>0$).

Starting in the first window ($i=1$), the relevant weights are $W_0$ and
$W_1$, which are initially set to zero. While simulating the first window,
we count the occurrence of states with $0 \leq S < S_{\rm M}$ ($f_{\rm
L}^1$) and $S_{\rm M} \leq S < \Delta S$ ($f_{\rm H}^1$), with $S_{\rm
M}=\Delta S/2$. To obtain the distribution in the number of particles
$\Pn$ (after all the quantity of interest) particle number histograms must
also be stored (note that $N$ fluctuates freely in each window). In the
first window, we thus record the probability distribution in the number of
particles $p_{\rm L}^1(N)$ for states with $0 \leq S < S_{\rm M}$, and
$p_{\rm H}^1(N)$ for states with $S_{\rm M} \leq S < \Delta S$. It is
recommended to store the distributions unnormalized. This makes it more
convenient to restart the simulations at a later stage in case higher
precision is required. After simulating the first window, the weight of
the higher window bound is updated to force a uniform distribution in $S$
using $W_{i, \rm new} = W_{i, \rm old} + \ln (f_{\rm H}^i / f_{\rm L}^i)$,
while keeping $W_{i-1}$ fixed, where $i$ is the window number.

In the second window ($i=2$), the relevant weights are $W_1$ and $W_2$. To
simulate efficiently, the weights $W_0$ and $W_1$ of the previous window
are extrapolated to estimate $W_2$. The extrapolated estimate is used
while simulating the second window, and the counts $f_{\rm L}^2$ and
$f_{\rm H}^2$ are recorded, as well as the distributions $p_{\rm L}^2(N)$
and $p_{\rm H}^2(N)$. After simulating the second window, $W_2$ is updated
as before, and the next window is considered.

The above procedure is repeated up to some maximum number of windows
$w_{\rm max}$ chosen well into the nematic phase. The remaining step is to
combine the weights $W_i$ with the distributions $p_{\rm L}^i(N)$ and
$p_{\rm H}^i(N)$ to obtain $\Pn$. Note that the upper region of window $i$
overlaps with the lower region of the next window $i+1$, see \fig{umb}.  
More precisely, the distributions $p_{\rm H}^i(N)$ and $p_{\rm
L}^{i+1}(N)$ stem from the same $S$ interval and are thus measured with
the same probability by the sampling scheme. Therefore, these
distributions may be combined $\bar{p}_i(N) = p_{\rm H}^i(N) + p_{\rm
L}^{i+1}(N)$, and normalized such that $\sum_{N=0}^\infty \bar{p}_i(N)=1$.
The distribution $\Pn$ is simply a weighted sum of the above (normalized)
$\bar{p}_i(N)$. Since $-\kb T \, W_i$ corresponds to a free energy, each
$\bar{p}_i(N)$ contributes to $\Pn$ with a weight proportional to
$e^{W_i}$. This leads to $\Pn = C \sum_{i=1}^{w_{\rm max}} \bar{p}_i(N)
e^{W_i}$, where the sum is over all windows, and normalization constant
$C^{-1} = \sum_{N=0}^\infty \sum_{i=1}^{w_{\rm max}} \bar{p}_i(N)
e^{W_i}$.

\subsection{Bias on $\rho$ versus bias on $S$}

Clearly, the discussed methods serve the same purpose: to measure the
distribution $\Pn$ at coexistence. Density biased sampling is by far the
easiest to implement. It has the additional advantage that the coexistence
chemical potential need not be specified beforehand: once $\Pn$ has been
measured at some chemical potential $\mu_0$, it can be extrapolated to any
other chemical potential $\mu_1$ by using the equation
\begin{equation}
\label{eq:ext}
  P(N|\mu_1) = P(N|\mu_0) e^{\beta (\mu_1 - \mu_0)N},
\end{equation}
with $P(N|\mu_\alpha)$ the probability distribution $\Pn$ at chemical
potential $\mu_\alpha$. Obviously, one should establish roughly beforehand
the density at which the IN transition occurs, to avoid sampling large
regions of irrelevant phase-space.

The situation is reversed when biasing on the nematic order parameter. In
this case, the sampling range is always from $S=0$ to $S \approx 1$.
However, to observe phase coexistence, it is essential to use a chemical
potential that is rather close to the coexistence value. Of course,
\eq{eq:ext} still holds, but the range in $\mu$ over which one can
extrapolate is much smaller, precisely because the bias is put on $S$ and
not on $\rho$. An estimate of the coexistence chemical potential may be
obtained in a density biased simulation of a small system, or via the
Widom insertion algorithm \cite{widom:1963, frenkel.smit:2001}. This
certainly makes biasing on $S$ more involved. Moreover, for each attempted
MC move, $S$ in the final state must be determined, regardless of whether
the move is accepted. It is therefore important to calculate $S$
efficiently. In particular, the ${\cal O}(N)$ loop of \eq{eq:s2} should be
eliminated, which can be done following the method outlined in
\olcite{vink.schilling:2005}.

\section{Results}
\label{sec:res}

An important conclusion of \olcite{vink.schilling:2005} is that the IN
interfacial tension obtained from $\Pn$ may be prone to strong finite size
effects. Away from any critical point, interfaces are the dominant source
of finite size effects. The use of periodic boundary conditions leads to
the formation of two interfaces. In small systems, the interfaces may
interact and this will influence the estimate of $\gin$. A convenient way
to suppress interface interactions is to use an elongated simulation box
with $\lz \gg \lx$ \cite{grossmann.laursen:1993}, in accord with
\fig{snap}. This forces an orientation of the interfaces perpendicular to
the elongated dimension (since this minimizes the interfacial area), with
a separation between the interfaces that is larger than it would be in a
cubic system of the same volume. The absence of interface interactions is
manifested by a pronounced flat region between the peaks in $\ln \Pn$.
Note that a flat region is essential, but not sufficient, to extract
$\gin$ reliably. There may still be finite size effects in the lateral
dimensions $\lx$ and $\ly$, arising for instance from capillary waves.
Ideally, the lateral dimensions should be large enough to capture the long
wavelength limiting form of the capillary spectrum
\cite{rowlinson.widom:1982}. To actually measure the capillary spectrum of
the IN interface is demanding \cite{akino.schmid.ea:2001}. A more
convenient approach sufficient for our purposes is to first establish a
minimum elongation $\lz$ in which interface interactions are suppressed,
and to then check for finite size effects in the lateral dimensions by
varying $\lx$ and $\ly$ explicitly.

An additional motivation to use large lateral dimensions is to stabilize
the interfaces. The interfacial free energy is of order $\gin \lx^2$, and
if this is small compared to $\kb T$, the interfaces will generally not be
stable. These issues are especially relevant for IN coexistence because
$\gin$ is very small. Therefore, in this section, we first perform MC
simulations in the canonical ensemble to obtain an indication of the
system size required to observe stable interfaces. Next, we present
coexistence data obtained using the nematic order biased sampling scheme.

\subsection{Interfacial profiles}
\label{sec:prof}

We consider hard rods, i.e.\ $\epsilon \to \infty$ in \eq{eq:pot}, of
elongation $L/D=15$. The simulations are performed in the canonical
ensemble, where the number of rods, the volume, and the temperature are
fixed. The box dimensions are $\lx=\ly=10 \, L/3$ and $\lz=20 \, L$. We
set the overall density of the system to $\rho^\star=0.205$, which is well
inside the coexistence region \cite{bolhuis.frenkel:1997,
dijkstra.roij.ea:2001}, corresponding to ca.~$11,000$ particles.  An
initial system is prepared containing two interfaces, with the director of
the nematic phase aligned in the plane of the interface. This is the
stable configuration, as confirmed by theory \cite{chen.noolandi:1992,
koch.harlen:1999} and simulation \cite{al-barwani.allen:2000,
mcdonald:2000, vink.schilling:2005}. The initial system is evolved with
random rotations and translations of single rods, accepted with the
standard Metropolis rules \cite{newman.barkema:1999, landau.binder:2000,
frenkel.smit:2001}. The system is equilibrated for $10^6$ sweeps, after
which a snapshot is taken every 260 sweeps, up to a total of $3 \times
10^4$ snapshots (one sweep corresponds to one attempted MC move per rod).

\begin{figure}
\begin{center}
\includegraphics[clip=,width=\columnwidth]{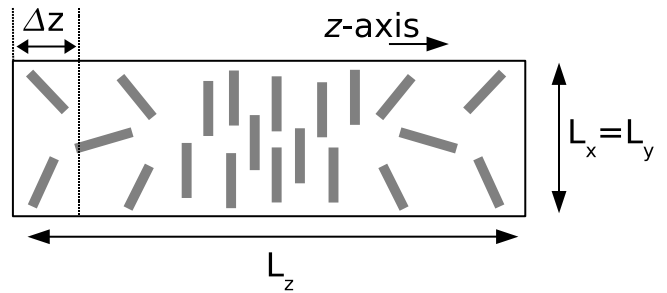}

\caption{\label{snap} Schematic representation of a simulation snapshot at IN
coexistence. The ordered nematic phase is located in the middle of the
simulation box. Profiles are measured along the $z$-dimension using bin size
$\Delta z$.}

\end{center}
\end{figure}

After equilibration, simulation snapshots schematically resemble \fig{snap}.
Note, however, that they contain far more particles than depicted in this simple
sketch. The aim is to measure the density profile $\rho(z)$, and the nematic
order parameter profile $S(z)$ along the elongated $z$-dimension, averaged over
many different snapshots. The averages are taken with the center of mass of the
snapshots shifted to the middle of the simulation box, with the constraint that
the nematic phase is also located in the middle, in accord with \fig{snap}. The
constraint is necessary to remove ambiguity arising from cases where the
isotropic phase is in the middle. Having shifted the center of mass, the density
profile is obtained by binning the $z$-axis in steps of $\Delta z \approx 0.17
L$. The local density $\rho(z)$ in a single snapshot is given by $n/v_{\rm B}$,
with $n$ the number of rods in the bin centered around $z$, and $v_{\rm B}$ the
volume of a single bin. The density profiles are then averaged over all
snapshots. Following \olcite{dijkstra.roij.ea:2001}, for the bin centered around
$z$ in a single snapshot, we also define a local orientational tensor $Q(z)$,
calculated using \eq{eq:s2} considering only the rods inside the bin. The local
orientational tensor elements are then averaged over all snapshots and $S(z) =
{\rm maxev} \, \avg{Q(z)}$.

\begin{figure}
\begin{center}

\includegraphics[clip=,width=\columnwidth]{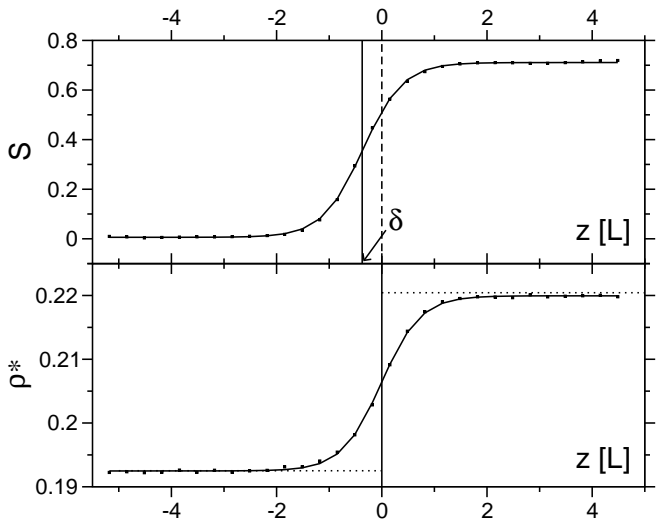} 

\caption{\label{profile} Nematic order parameter profile $S(z)$ (top), and
density profile (bottom) for hard rods with $L/D=15$ across the IN
interface. The shift between the profiles is marked $\delta$. Points are
raw simulation data; curves are hyperbolic tangent fits. The horizontal
lines in the lower frame represent the bulk isotropic and nematic
densities obtained in the grand canonical simulations of \sect{sec:hr}.}

\end{center}
\end{figure}

The averaged profiles are shown in \fig{profile}. The solid curves are
hyperbolic tangent fits of the form $A + B \tanh \left( \frac{z-z_c}{w}
\right)$, which describe the data well. Note that the profiles are shifted
with respect to each other. The magnitude of the shift, measured between
the inflection points, equals $\delta = 0.37 \pm 0.04 \, L$. This is
consistent with theoretical predictions $\delta = 0.45-0.5 \, L$
\cite{chen.noolandi:1992, shundyak.roij:2003}, as well as $\delta \approx
0.33 \, L$ obtained in simulations of ellipsoids \cite{allen:2000}. Note
that the simulated profiles are broadened due to capillary waves
\cite{akino.schmid.ea:2001}. Moreover, we observed considerable
fluctuations in the amount of isotropic and nematic phase during the
simulation, leading to large fluctuations in the interface positions along
the elongated $\lz$ dimension. The width of the averaged profile obtained
by fixing the center of mass is therefore additionally broadened
\cite{binder.muller:2000, tepper.briels:2002}. Because of these effects,
we cannot compare the interfacial width of the simulated profiles to
theoretical predictions. More important for our purposes, however, is the
observation that the interfaces are stable. For hard rods, the current
system size thus seems sufficient to accommodate stable interfaces.
 
\subsection{Comparison of $\rho$ and $S$ biased sampling}

\begin{figure}
\begin{center}

\includegraphics[clip=,width=\figwidth]{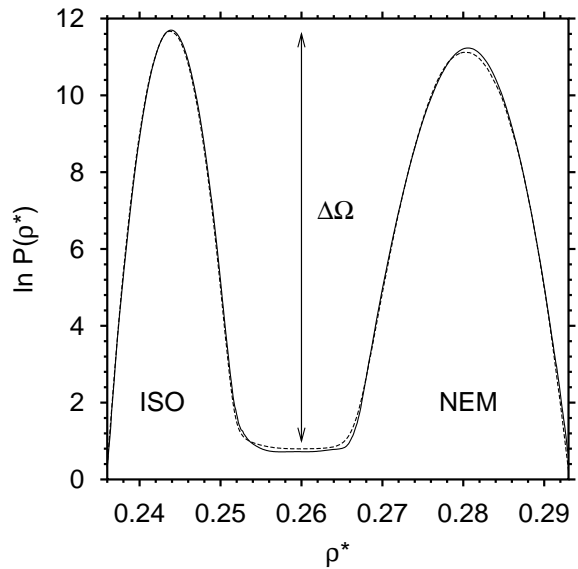}

\caption{\label{compare} Coexistence distribution $P(\rho^\star)$ for soft
rods with $L/D=15$ and $\beta\epsilon=2$ using box dimensions $\lx=\ly=2.1
\, L$ and $\lz=8.4 \, L$. The solid curve was obtained using a bias on the
nematic order parameter $S$; the dashed curve by using a bias on the
density. The average peak height $\Delta\Omega$, multiplied by $\kb T$,
equals the free energy barrier separating the isotropic (ISO) from the
nematic (NEM) phase.}

\end{center}
\end{figure}

Having established the typical system size required to observe stable
interfaces, biased sampling on the nematic order parameter is considered
next. First, we show that density and nematic order biased sampling yield
the same distribution $\Pn$. To this end, we consider a small system of
soft rods with $L/D=15$ and $\beta\epsilon=2$, in a simulation box of size
$\lx=\ly=2.1 \, L$ and $\lz=8.4 \, L$. The latter system was investigated
in previous work using density biased sampling \cite{vink.schilling:2005}.
The corresponding coexistence chemical potential reads as $\beta\mu
\approx 5.15$. The nematic order biased sampling scheme is applied to the
same system using the latter chemical potential and $\Delta S=0.002$, see
\fig{compare}. Shown is the coexistence distribution $\Pn$ obtained using
a bias on $S$ (solid curve), as well as using a bias on density (dashed
curve, reproduced from \olcite{vink.schilling:2005}). The agreement
between both methods is strikingly confirmed, thereby justifying the
approach of \sect{sec:s2}. For small systems, the required CPU time is
roughly equal for both methods. The data sets of \fig{compare} required
ca.~700 CPU hours each, on 2.2 GHz Pentium machines.

\subsection{Interfacial tension of soft rods}

\begin{figure}
\begin{center}

\includegraphics[clip=,width=\figwidth]{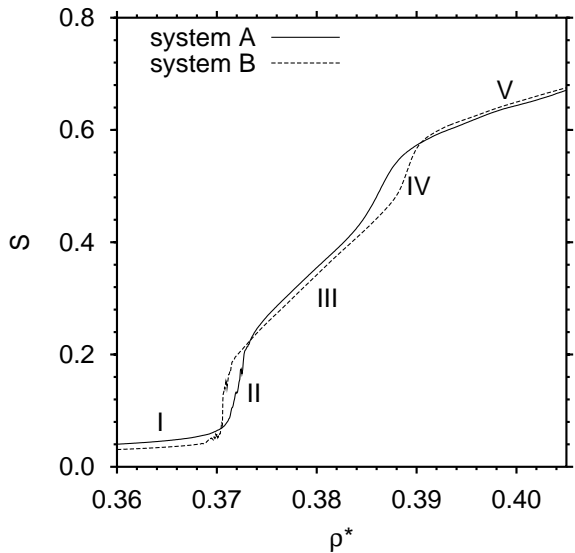}

\caption{\label{path} Dependence of the nematic order parameter on the
density for soft rods with $L/D=10$ and $\beta\epsilon=2$, obtained using
two different system sizes.}

\end{center}
\end{figure}

\begin{figure}
\begin{center}

\includegraphics[clip=,width=\figwidth]{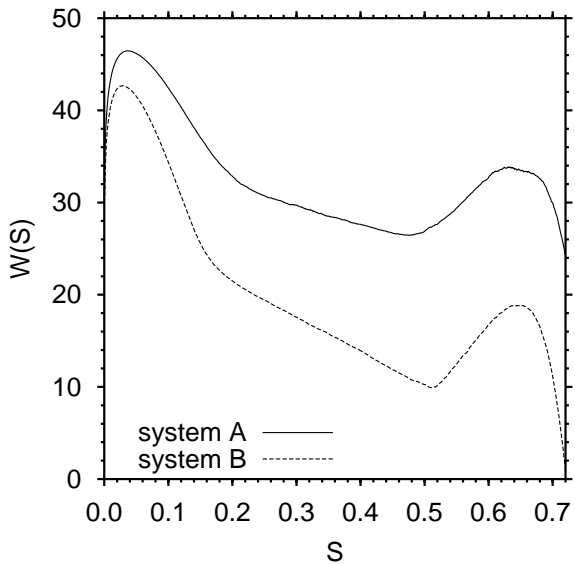}

\caption{\label{ps2} Weight function $W(S)$ obtained by biasing on the
nematic order parameter $S$ (see \sect{sec:s2}) for soft rods with
$L/D=10$ and $\beta\epsilon=2$ using two different system sizes.}

\end{center}
\end{figure}

\begin{figure}
\begin{center}

\includegraphics[clip=,width=\figwidth]{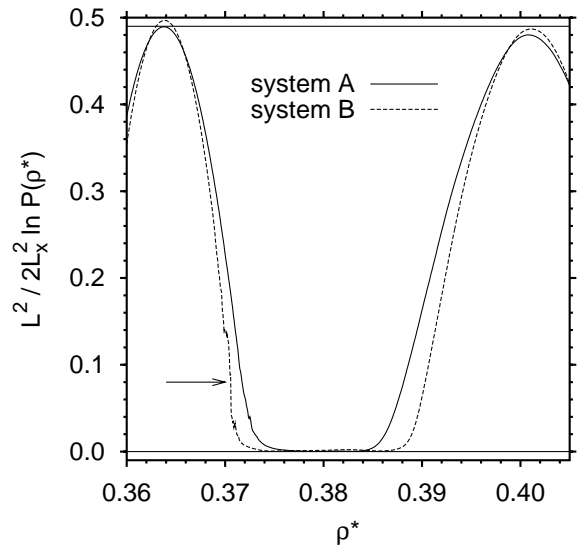}

\caption{\label{large} Logarithm of $P(\rho^\star)$ at coexistence for
soft rods with $L/D=10$ and $\beta\epsilon=2$ using two different system
sizes. Note the flat region in between the peaks. The arrow indicates
transition~II of \fig{path}.}

\end{center}
\end{figure}

Next, we consider soft rods with $L/D=10$ and $\beta\epsilon=2$. We aim to
accurately measure the interfacial tension. To this end, large system
sizes are required such that a bias on $S$ is essential. As explained
before, the elongated $\lz$ dimension of the simulation box must be large
enough to accommodate non-interacting interfaces. At the same time, $\lx$
and $\ly$ must be large enough to suppress finite size effects in the
lateral dimensions. We therefore consider two system sizes: $\lx=\ly=3.5
\, L; \, \lz=10.5 \, L$ (system A), and $\lx=\ly=4 \, L; \, \lz=14 \, L$
(system B), where the lateral dimensions are deliberately chosen to exceed
those of \sect{sec:prof}. The simulations are performed using $\Delta
S=0.001$ and $0.002$, for system A and B, respectively. An initial
estimate of the coexistence chemical potential was taken from previous
work \cite{vink.schilling:2005}.

In \fig{path}, the dependence of the nematic order parameter on the
number of particle is shown, calculated using 
\begin{equation}
\label{eq:s2avg}
  S(N) = C \sum_{i=1}^{w_{\rm max}} S_i \bar{p}_i(N) e^{W_i}, 
\end{equation}
with $S_i = i \Delta S/2 - \Delta S/4$, and the remaining symbols defined
as before. Analogous to fluid-vapor transitions
\cite{virnau.muller.ea:2004, virnau.macdowell.ea:2004,
virnau.macdowell.ea:2003}, five distinct regions can be distinguished. In
region~I, a single isotropic phase is observed. Region~II corresponds to
the transition from the bulk isotropic phase, to the phase with two
parallel interfaces. The transition is characterized by the formation of a
nematic droplet in an isotropic background, which grows with the density
until it self-interacts through the periodic boundaries, ultimately
leading to two parallel interfaces. In region~III, the interfaces have
formed and the system is at coexistence, schematically resembling
\fig{snap}. Increasing the density further leads to a growth of the
nematic domain, at the expense of the isotropic domain. Region~IV
corresponds to the transition to the pure nematic phase, during which the
system is characterized by an isotropic droplet in a nematic background.
In region~V, finally, a single nematic phase is observed.

In \fig{ps2}, we show the corresponding weight function $W(S)$ for both
systems. The double-peaked structure is clearly visible. Note that the
isotropic peak is significantly higher than the nematic peak. This
indicates that the chemical potential used in the simulations is below the
coexistence value. Since coexistence is defined by equal weight in the
peaks of $\Pn$, and not in $W(S)$, \fig{ps2} cannot be used to obtain the
coexistence chemical potential. Instead, $\Pn$ must be constructed first,
by combining $W(S)$ with the single window distributions $\bar{p}_i(N)$;
\eq{eq:ext} may then be used to extrapolate $\Pn$ to coexistence. The
resulting coexistence chemical potential equals $\beta\mu \approx 7.13$
for both systems. In \fig{large}, the logarithm of $P(\rho^\star)$ at
coexistence is plotted for both systems, scaled with $L^2/(2\lx^2)$, and
the plateaus shifted to zero. In this way, the barrier directly reflects
the interfacial tension $\gin$, in units of $\kb T/L^2$
\cite{binder:1982}. An important observation is that the peaks in both
distributions are separated by a pronounced flat region. This shows that
the elongated $\lz$ dimension of the simulation box is sufficient.
Moreover, the peak heights are similar, indicating that finite size
effects in the lateral dimensions $\lx$ and $\ly$ are also small.
Therefore, we conclude that the barrier in \fig{large} accurately reflects
the interfacial tension $\gin$ for soft rods with $L/D=10$ and
$\beta\epsilon=2$. The resulting estimate reads as $\gin = 0.49 \, \kb
T/L^2 = 0.0049 \, \kb T/D^2$.

In the nematic phase, ca.~6000 rods were simulated for system A, and
10,000 for system B. To obtain reliable results, a substantial investment
in CPU time is thus required (ca.~3200 CPU hours were invested for
system~B). Since biased sampling schemes are easy to parallelize, results
can typically be obtained within 1-2 weeks on a modern computer cluster.
Accurate sampling is especially important around transitions II and IV,
and this becomes increasingly difficult in large systems
\cite{virnau.macdowell.ea:2003}. This may already be inferred from the
scatter in the data of system~B around transition~II (arrow in
\fig{large}). Transition~IV, on the other hand, is sampled with
surprisingly little difficulty. The likely explanation is that process~II
requires the formation of a nematic nucleus whose director is aligned in
the $xy$-plane. Process~IV, on the other hand, does not require any
preferred orientation of the (isotropic) nucleus, and is therefore easier
to sample.

\subsection{Consequences for finite size extrapolation}

\begin{figure}
\begin{center}

\includegraphics[clip=,width=\figwidth]{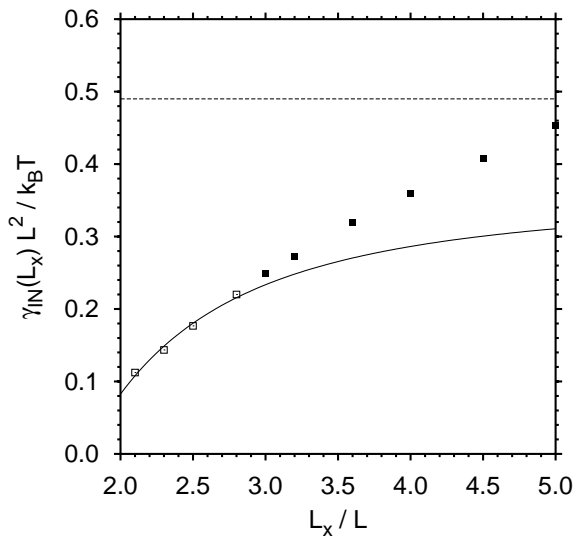}

\caption{\label{fss} Interfacial tension $\gin(\lx)$ obtained in cubic
systems with edge $\lx$, as function of $\lx$, for soft rods with $L/D=10$
and $\beta\epsilon=2$.}

\end{center}
\end{figure}

An alternative method to obtain the interfacial tension is to measure
$\gin(\lx)$ in cubic systems with edge $\lx$, and use the extrapolation
equation of Binder \cite{binder:1982}
\begin{equation}
\label{eq:fss}
  \gin(\lx) = \gin + a/\lx^2 + b \ln(\lx)/\lx^2, 
\end{equation}
to estimate $\gin$. In principle, this approach enables estimates of
$\gin$ through an elimination of finite size effects, but it requires
estimates over a range of values for which $\gin(\lx) \lx^2 / \kb T \gg
1$. In practice, however, one often tries to use \eq{eq:fss} using data
from smaller systems. In \olcite{vink.schilling:2005}, this approach was
applied to soft rods with $L/D=10$ and $\beta\epsilon=2$, assuming $b=0$
in \eq{eq:fss}, leading to $\gin = 0.0035 \, \kb T/D^2$. This estimate
differs profoundly from the one of the previous section, implying that 
finite size extrapolation must be used with care. The issue is
investigated further in \fig{fss}. Shown is $\gin(\lx)$ as function of
$\lx$, where the open squares are data from \olcite{vink.schilling:2005},
and closed squares data from larger systems obtained in this work. The
horizontal line corresponds to the estimate of \fig{large}. Note that the
data indeed approach the latter estimate. The curve is a fit to the open
squares using \eq{eq:fss} with $b=0$, which summarizes the result of
\olcite{vink.schilling:2005}. Clearly, the fit fails to capture the data
of the larger systems. Allowing $b$ in \eq{eq:fss} to be non-zero will
obviously lead to a better fit, but the resulting $\gin$ depends
sensitively on the range over which the fit is performed, making this
approach somewhat arbitrary. The problem partly stems from the difficulty
in distinguishing $a/\lx^2$ numerically from $b\ln(\lx)/\lx^2$, since the
range in $\lx$ that can be sampled is rather small. Additionally, in small
systems, the interface interactions may be strong. This will introduce
corrections to \eq{eq:fss}, which may even yield non-monotonic behavior in
$\gin(\lx)$ \cite{mon:1988, berg.hansmann.ea:1993*b,
hunter.reinhardt:1995}. As a result, it is difficult to extract $\gin$ via
finite size extrapolation. In contrast, by using an elongated simulation 
box, and by explicitly checking for finite size effects in the lateral
dimensions, $\gin$ can be extracted reliably as shown in \fig{large}.
This, we conclude in hindsight, should be the method of choice.

\subsection{Interfacial tension of hard rods}
\label{sec:hr}

\begin{figure}
\begin{center}

\includegraphics[clip=,width=\figwidth]{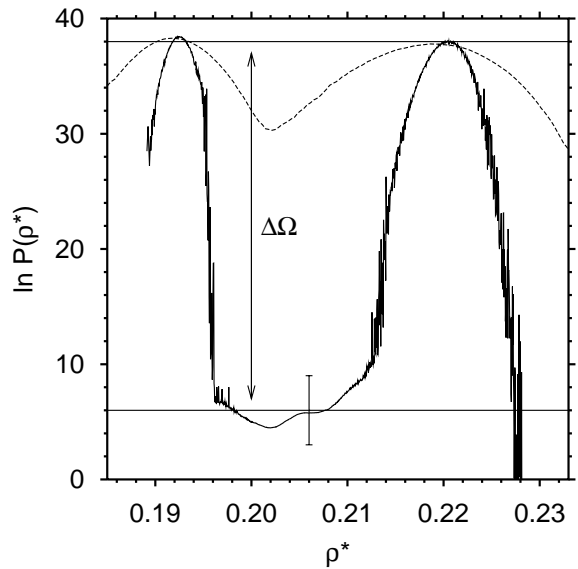}

\caption{\label{hard} Coexistence distribution (solid curve) for hard rods
with $L/D=15$, obtained using box dimensions $\lx=\ly=10 \, L/3$ and
$\lz=10 \, L$. The barrier $\Delta\Omega$ is measured between the
horizontal lines, where the bar gives an indication of the uncertainty.
The dashed curve shows $P(\rho^\star)$ obtained in a smaller cubic box
with $\lx=2.3 \, L$.}

\end{center}
\end{figure}

Finally, we apply nematic order biased sampling to a system of hard rods
with $L/D=15$, system size $\lx=\ly=10 \, L/3$ and $\lz=10 \, L$,
corresponding to ca.~6000 rods in the nematic phase. An initial estimate
of the coexistence chemical potential was obtained via Widom insertion
\cite{widom:1963}. The nematic order parameter is sampled with resolution
$\Delta S=0.0025$ to obtain $W(S)$. Combining $W(S)$ with the single
window distributions $\bar{p}_i(N)$ and applying \eq{eq:ext} yields for
the coexistence chemical potential $\beta\mu \approx 5.58$. The
corresponding coexistence distribution is shown in \fig{hard}.

Note that $\Pn$ for hard rods is prone to substantial statistical error.  
This is to be expected because the acceptance rate of grand canonical
insertion for hard rods is only 0.004\%, compared to 8\% for soft rods.
Nevertheless, the double-peaked structure is clearly visible. From the
average peak locations, we obtain $\riso=0.193$ and $\rnem=0.220$. The
latter densities are consistent with the bulk plateaus in the density
profile, indicated by the horizontal lines in \fig{profile}. To further
check the consistency of our results, an additional simulation in a
smaller cubic system with $\lx=2.3 \, L$ was performed; the corresponding
coexistence distribution is shown dashed in \fig{hard}. Of course, this
system is too small to extract the interfacial tension, but the peak
positions, and hence the coexistence densities, agree well with those of
the larger system. The agreement with bulk densities obtained via Gibbs
ensemble simulations \cite{dijkstra.roij.ea:2001} and Gibbs-Duhem
integration \cite{bolhuis.frenkel:1997} is better than 4\%. The height of
the free energy barrier of the larger system reads as $\Delta \Omega = 32
\pm 3 \, \kb T$, leading to an interfacial tension $\gin \approx 1.4 \,
\kb T/L^2 = 0.0064 \, \kb T/D^2 = 0.096 \, \kb T/LD = 0.10 \, \kb T /
(L+D)D$.

\section{Discussion and summary}

In this paper, we have presented methodic developments that allow for the
estimation of the interfacial tension between isotropic and nematic phases
in suspensions of rods. The problem is challenging because $\gin$ is very
small, and methods that work well for interfaces between isotropic phases
become problematic, such as exploiting the anisotropy of the pressure
tensor \cite{rowlinson.widom:1982}, or analyzing the capillary wave
spectrum (the latter requires very precise data from huge systems
\cite{akino.schmid.ea:2001}). The novelty of the present approach is to
combine grand canonical MC simulations with a bias on the nematic order
parameter, and obtain $\gin$ from the grand canonical distribution $\Pn$.
The advantage is that the problem of ``jamming'' is largely solved,
enabling simulations of large systems.

\begin{figure}
\begin{center}

\includegraphics[clip=,width=\figwidth]{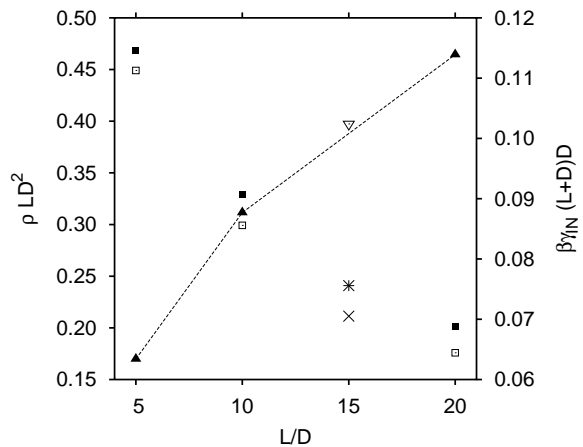}

\caption{\label{velasco} IN coexistence properties of the hard-rod fluid
obtained by theory \cite{velasco.mederos:2002}, compared to simulation
results obtained in this work. Symbols are explained in the text.}

\end{center}
\end{figure}

The current approach also allows for grand canonical simulations of hard
rods, enabling a direct comparison to theory. In the Onsager limit of
infinite rod length, theoretical estimates of $\gin$ typically range from
0.156 \cite{shundyak.roij:2001} to 0.34 \cite{mcmullen:1988}, in units of
$\kb T / LD$. As expected, this exceeds the value for hard rods obtained
in this work ($\gin \approx 0.096 \, \kb T/LD$)  because $L/D=15$ is still
far from the Onsager limit. As shown by experiment \cite{chen.gray:2002}
and theory \cite{velasco.mederos:2002}, $\gin$ increases with $L/D$. The
latter theory is based on the Somoza-Tarazona density functional
\cite{somoza.tarazona:1989} and its main findings are summarized in
\fig{velasco}. Shown are the coexistence densities (left axis) and the
interfacial tension (right axis) as function of the rod elongation $L/D$,
where we have adopted the units of \olcite{velasco.mederos:2002}. Open and
closed squares show the theoretical density of the isotropic and the
nematic phase, respectively;  the star and the cross are the corresponding
simulation estimates of this work. Closed triangles are the theoretical
interfacial tension, where the line serves to guide the eye; the open
triangle represents the simulation estimate of $\gin$. Theoretical
estimates are reported for $L/D=5,10,20$, but unfortunately not for
$L/D=15$. This makes a direct comparison difficult; interpolation of the
theoretical results, however, seems in good agreement with our simulation
results, as may be inferred from \fig{velasco}.

A typical rod dimension in experiments is $L=150$~nm and $\gin =
0.00083$~mN/m \cite{chen.gray:2002}. For $T=298$~K, this length translates
into $0.00025$~mN/m using our estimate of $\gin$. Obviously, this estimate
differs from the experimental one because the hard-rod fluid is a
simplified model, but it is reassuring to see that the order of magnitude
is confirmed.

The current biased sampling scheme thus seems well suited to simulate IN
coexistence, even for hard interactions. Our scheme may also be useful for
the application of transition path sampling
\cite{bolhuis.chandler.ea:2002} to anisotropic colloidal systems, since it
can provide valuable starting paths; work along these lines is in
progress. The remaining bottleneck is the low acceptance rate of grand
canonical insertion. It remains a challenge to address this final problem.
Since the overall density around the IN transition is low, it is
anticipated that higher acceptance rates can be realized using smarter
insertion schemes. To develop such schemes would be the subject of future
work.

\acknowledgments

We thank the Deutsche Forschungsgemeinschaft (DFG) for support (TR6/A5 and
TR6/D5) and K. Binder, M. M\"uller, P. Virnau, P. van der Schoot, R. van
Roij and M.  Dijkstra for a careful reading of the manuscript and/or
helpful suggestions. TS is supported by the Emmy Noether program of the
DFG. Allocation of computer time on the JUMP at the Forschungszentrum
J\"ulich is gratefully acknowledged.

\bibstyle{revtex}
\bibliography{mainz}

\end{document}